%% file: belle.tex
\RequirePackage{xspace}
\documentclass[aps,reprint,tightenlines,superscriptaddress]{revtex4}
\usepackage{graphicx}
\usepackage{dcolumn}
\usepackage{bm}
\usepackage{multirow}
\usepackage{amsmath}
\usepackage[title]{appendix}

\def\ep         {\ensuremath{e^+}\xspace}
 
\def\epem       {\ensuremath{e^+e^-}\xspace}

\def\mumu       {\ensuremath{\mu^+\mu^-}\xspace}

\def\ellell     {\ensuremath{\ell^+ \ell^-}\xspace}

\def\g     {\ensuremath{\gamma}\xspace}

\def\qqbar {\ensuremath{q\overline q}\xspace}
\def\QQbar {\ensuremath{Q\overline Q}\xspace}

\def\bbbar {\ensuremath{b\overline b}\xspace}

\def\pizz   {\ensuremath{\pi^0}\xspace}

\def\pipi  {\ensuremath{\pi^+\pi^-}\xspace}

\def\om   {\ensuremath{\omega}\xspace}

\def\Kbar  {\kern 0.2em\overline{\kern -0.2em K}{}\xspace}

\def\Kz    {\ensuremath{K^0}\xspace}
\def\Kzb   {\ensuremath{\Kbar^0}\xspace}
\def\KzKzb {\ensuremath{\Kz \kern -0.16em \Kzb}\xspace}
\def\Kp    {\ensuremath{K^+}\xspace}
\def\Km    {\ensuremath{K^-}\xspace}

\def\KpKm  {\ensuremath{\Kp \kern -0.16em \Km}\xspace}

\def\Dbar    {\kern 0.2em\overline{\kern -0.2em D}{}\xspace}

\def\Dz      {\ensuremath{D^0}\xspace}
\def\Dzb     {\ensuremath{\Dbar^0}\xspace}
\def\DzDzb   {\ensuremath{\Dz {\kern -0.16em \Dzb}}\xspace}
\def\Dp      {\ensuremath{D^+}\xspace}
\def\Dm      {\ensuremath{D^-}\xspace}

\def\DpDm    {\ensuremath{\Dp {\kern -0.16em \Dm}}\xspace}

\def\Bbar    {\kern 0.18em\overline{\kern -0.18em B}{}\xspace}

\def\Bz      {\ensuremath{B^0}\xspace}
\def\Bzb     {\ensuremath{\Bbar^0}\xspace}
\def\BzBzb   {\ensuremath{\Bz {\kern -0.16em \Bzb}}\xspace}
\def\Bu      {\ensuremath{B^+}\xspace}
\def\Bub     {\ensuremath{B^-}\xspace}

\def\BpBm    {\ensuremath{\Bu {\kern -0.16em \Bub}}\xspace}

\def\jpsi     {\ensuremath{{J\mskip -3mu/\mskip -2mu\psi\mskip 2mu}}\xspace}

\def\cb {\ensuremath{\chi_{bJ}}\xspace}
\def\cbn    {\ensuremath{\chi_{bJ}(nP)}\xspace}
\def\cbj#1  {\ensuremath{\chi_{bJ}{(#1P)}}\xspace}

\def\chibOne#1  {\ensuremath{\chi_{b1}{(#1P)}}\xspace}
\def\chibTwo#1  {\ensuremath{\chi_{b2}{(#1P)}}\xspace}
\def\chibZero#1  {\ensuremath{\chi_{b0}{(#1P)}}\xspace}

\def\chib12   {\ensuremath{\chi_{b1,2}{(2P)}}\xspace}
\def\Y#1S{\ensuremath{\Upsilon{(#1S)}}\xspace}% no space before {...}!
\def\OneS  {\Y1S}
\def\TwoS  {\Y2S}
\def\ThreeS{\Y3S}
\def\FourS {\Y4S}

\def\dmchi{\ensuremath{\Delta M_{\chi}}\xspace}
\def\dmpp{\ensuremath{\Delta M_{\pi\pi}}\xspace}
\def\mom{\ensuremath{M_{\omega}}\xspace}

\mathchardef\Deltares="7101
\def\Deltabar{\kern 0.25em\overline{\kern -0.25em \Deltares}{}\xspace}
\def\Lbar{\kern 0.2em\overline{\kern -0.2em\Lambda\kern 0.05em}\kern-0.05em{}\xspace}
\def\Sigbar{\kern 0.2em\overline{\kern -0.2em \Sigma}{}\xspace}
\def\Xibar{\kern 0.2em\overline{\kern -0.2em \Xi}{}\xspace}
\def\Obar{\kern 0.2em\overline{\kern -0.2em \Omega}{}\xspace}
\def\Nbar{\kern 0.2em\overline{\kern -0.2em N}{}\xspace}
\def\Xb{\kern 0.2em\overline{\kern -0.2em X}{}\xspace}

\newcommand{\tev}{\ensuremath{\mathrm{\,Te\kern -0.1em V}}\xspace}
\newcommand{\gev}{\ensuremath{\mathrm{\,Ge\kern -0.1em V}}\xspace}
\newcommand{\mev}{\ensuremath{\mathrm{\,Me\kern -0.1em V}}\xspace}
\newcommand{\kev}{\ensuremath{\mathrm{\,ke\kern -0.1em V}}\xspace}
\newcommand{\ev}{\ensuremath{\mathrm{\,e\kern -0.1em V}}\xspace}
\newcommand{\gevc}{\ensuremath{{\mathrm{\,Ge\kern -0.1em V\!/}c}}\xspace}
\newcommand{\mevc}{\ensuremath{{\mathrm{\,Me\kern -0.1em V\!/}c}}\xspace}
\newcommand{\gevcc}{\ensuremath{{\mathrm{\,Ge\kern -0.1em V\!/}c^2}}\xspace}
\newcommand{\mevcc}{\ensuremath{{\mathrm{\,Me\kern -0.1em V\!/}c^2}}\xspace}

\def\cm   {\ensuremath{{\rm \,cm}}\xspace}

\def\invfb   {\ensuremath{\mbox{\,fb}^{-1}}\xspace}

\def\mus  {\ensuremath{\rm \,\mus}\xspace}

\def\mus        {\ensuremath{\,\mu{\rm s}}\xspace}    %% microsecond
\def\to                 {\ensuremath{\rightarrow}\xspace}

\def\gsim{{~\raise.15em\hbox{$>$}\kern-.85em
          \lower.35em\hbox{$\sim$}~}\xspace}
\def\lsim{{~\raise.15em\hbox{$<$}\kern-.85em
          \lower.35em\hbox{$\sim$}~}\xspace}

\def \asy#1#2{\ensuremath{^{+#1}_{-#2}}\xspace}

\def \j#1{\ensuremath{J = #1}\xspace}

\def \evtgen{{\sc E{\scriptsize VT}G{\scriptsize EN}}}
\def \photos{{\sc P{\scriptsize HOTOS}}}
\def \geant3{{\sc GEANT3}}

\begin{document}

\makebox[0.98\textwidth][r]{BELLE-CONF-2102}

\title{ \quad\\[0.5cm]  Study of $\bm{\chi_{bJ}(nP)\rightarrow\omega\Upsilon(1S)}$ at Belle}
\input{authors}

\begin{abstract}

We report results from a study of hadronic transitions of the $\chi_{bJ}(nP)$ states of bottomonium at Belle. The $P$-wave states are reconstructed in transitions to the $\Upsilon(1S)$ with the emission of an $\omega$ meson. The transitions of the $n=2$ triplet states provide a unique laboratory in which to study nonrelativistic quantum chromodynamics, as the kinematic threshold for production of an $\omega$ and $\Upsilon(1S)$ lies between the \j0 and \j1 states. A search for the $\chi_{bJ}(3P)$ states is also reported.

\end{abstract}

\maketitle
\tighten

\section{Introduction} \label{sec:introduction}

Recently, the hadronic transitions among heavy quarkonium $(\QQbar,~\text{where}~Q=c,b)$ states have been the focus of detailed study \cite{belle_5Sdipi,babar_Y4S_eta_Y1S,babar_x3872_omegaJpsi,babar_YnS_eta_Y1S,belle_hbDiscovery,4sEta_hb,guidoEtaDiPi,guidoEtaPrime,belle_YmS_omega_etab,bes3_x3872_omegaJpsi}. 
Of such transitions, those occurring near kinematic thresholds for the decay constitute a unique laboratory in which to study the emission and hadronization of soft gluons \cite{Brambilla:quarkoniumReview}.

The recent observation of the near-threshold transition $\chi_{c1}(3872)\to\om\jpsi$ by BESIII \cite{bes3_x3872_omegaJpsi} is of particular interest. Although it is a narrow state ($\Gamma_{\chi_{c1}(3872)} = 1.19\pm0.21$~\mev~\cite{pdg}) that lies nearly $8~\mev$ below the kinematic threshold for production of an \om and \jpsi meson, the observed branching is as large as the discovery channel $\chi_{c1}(3872)\to\pipi\jpsi$, with a relative branching ratio of $1.1\pm0.4$ \cite{Belle:x3872_discovery,pdg}. In the bottomonium $(\bbbar)$ sector, the analogous $\om\OneS$ final-state threshold lies between the \j0 and \j1 states of the \cbj2 triplet. 

In 2004, CLEO reported the first observation of the transitions $\cbj2 \to\om\OneS$ produced in radiative decays of $(5.81\pm0.12)\times10^6~\ThreeS$ mesons\cite{toddOmega}. The branching fractions of the \j1 and \j2 states were measured to be on the order of $1\%$. Since their discovery, no confirmation of the branching fraction measurements has been made. Although no indication of a sub-threshold \j0 signal was seen by CLEO due to limited statistics, Monte Carlo (MC) simulation of \chibZero2 transitions to an $S$-wave $\om\OneS$ indicates that the decay may be observed, though in such transitions the \om lineshape is distorted due to the presence of the nearby kinematic threshold.

In this paper, we report an inclusive search for the bottomonium states \cbj2 and \cbj3 produced in radiative transitions of the \ThreeS and \FourS. The hadronic transitions $\cbn \to\om\OneS$ are studied by reconstructing $\OneS\to\ellell$ with $\ell=e,\mu$. The \om meson is reconstructed in its decay to $\pipi\pizz$, with $\pizz\to\g\g$. We measure the branching fractions of the hadronic transitions along with the cascade branching ratio,
\begin{equation}
    \label{eqn:rJ1}
    r_{J/1} = \frac{\mathcal{B}\left( \ThreeS \to\g\cbj2 \to\g\om\OneS \right)}{\mathcal{B}\left( \ThreeS \to\g\chibOne2 \to\g\om\OneS \right)},
\end{equation}
and compare with the expectation from the QCD multipole expansion (QCDME) model \cite{Voloshin:omegaDecayComments}, which we calculate using the current world averages \cite{pdg}. 

As no significant \cbj3 signal is seen, we set an upper limit on the dominant cascade branching fraction $\mathcal{B}\left(\FourS\to\g\chibOne3 \to\g\om\OneS\right)$.

\section{Data Samples and Detector} \label{sec:data_samples_and_detector}

We analyze data samples corresponding to an integrated luminosity of $3~\invfb$ and $513~\invfb$ accumulated near the \ThreeS and \FourS resonances, respectively, by the Belle detector \cite{BelleDetector} at the KEKB asymmetric-energy \epem collider \cite{KEKB}. We also study a sample, referred to as the off-resonance sample, collected about 60 MeV below the \FourS resonance, totalling $56~\invfb$. The population of \ThreeS in the combined dataset is determined from a reconstruction of $\ThreeS\to\pipi\OneS[\ellell]$ to be $(27.9\pm1.0)\times10^6$ mesons (See Appendix \ref{app:3S_count} for details). Decays of \ThreeS mesons in data accumulated above the \ThreeS resonance are assumed to come from initial state radiation (ISR) by the \epem pair. No attempt has been made to reconstruct the ISR photons, which typically emerges at small angles to the beampipe \cite{benayoun1999ISR}.

The Belle detector is a large-solid-angle magnetic
spectrometer that consists of a silicon vertex detector (SVD),
a 50-layer central drift chamber (CDC), an array of
aerogel threshold Cherenkov counters (ACC), 
a barrel-like arrangement of time-of-flight
scintillation counters, and an electromagnetic calorimeter
(ECL) comprised of CsI(Tl) crystals located inside 
a superconducting solenoid coil that provides a 1.5~T
magnetic field. The ECL is divided into three regions spanning the angle of inclination $(\theta)$ with respect to the direction opposite the \ep beam (taken to be the $z$ axis). The ECL backward endcap, barrel, and forward endcap, cover ranges $\cos\theta \in [-0.91,-0.65]$, $\cos\theta \in [-0.63,0.85]$, and $\cos\theta\in[0.85,0.98]$, respectively. An iron flux-return located outside of the coil is instrumented with resistive plate counters to detect $K_L^0$ mesons and to identify muons (KLM). The data collected for this analysis used an inner detector system with a 1.5~\cm beampipe, 
a 4-layer SVD and a small-cell inner drift chamber. The detector is described in detail elsewhere~\cite{BelleDetector}.

A set of event selection criteria are devised to optimize the retention of signal events while suppressing backgrounds from mis-reconstructed \pizz decays, resonant \bbbar decays, and non-resonant (continuum) production of other quark and lepton species. For all optimizations, we employ a figure of merit FOM$\equiv S/\sqrt{S+B}$, where $S$ and $B$ denote the number of signal and background events, respectively. To study these criteria and their associated reconstruction efficiencies, MC simulated events are produced. MC events are generated with the \evtgen \cite{evtgen} package. Radiative transitions among \bbbar states are generated with the helicity-amplitude formalism \cite{HELAMP}. Di-pion transitions among the \ThreeS, \TwoS, and \OneS states produced according to the matrix elements reported in Ref. \cite{cleoDipionAngularAnalysis}. All other di-pion transitions as well as the hadronic transitions $\cbn \to\om\OneS$ are modeled with phase space. The decay of the \om meson is simulated uniformly across the Dalitz plot. Final state radiation effects are modeled by \photos~\cite{PHOTOS}. The Belle detector response is simulated with GEANT3 \cite{geant3}.

\section{Event Selection} \label{sec:event_selection}

Slight differences exist in the event selection criteria depending on the dataset and decay channel. Where appropriate, these differences are labeled according to the dataset and radial quantum number $(n)$ of the \cbn triplet. Charged tracks are required to originate within $2.0~\cm$ of the interaction point (IP) along the $z$ axis and within $0.5~\cm$ in the transverse plane. Tracks whose momentum exceeds $4~\gev$ measured in the center-of-mass (CM) frame are preliminarily identified as leptons, and pairs of such tracks are combined to form \OneS candidates if their invariant mass lies within the range $M(\ellell)\in[9.0,9.8]~\gev$.  A fiducial selection is made by requiring that an event contains only one \OneS candidate.

A likelihood $\mathcal{L}_i~(i=\mu,\pi,K)$ is ascribed to each charged track based on its signature in the KLM and its agreement with an extrapolation of the track from the CDC. The muon identification likelihood ratio is defined as\\ $\mathcal{R}_\mu = \mathcal{L}_\mu / (\mathcal{L}_\mu + \mathcal{L}_\pi + \mathcal{L}_K)$. A similar electron identification likelihood ratio $\mathcal{R}_e$ is constructed for electrons using measurements from the CDC, ECL, and ACC \cite{bellePID}. Both lepton candidates are identified as muons if $\mathcal{R}_\mu > \mathcal{R}_e$ for either candidate; otherwise, they are considered as electrons. 

QCD and QED continuum processes of the form $\epem\to\qqbar$, where $q=u,d,s,c$, and $\epem\to(\epem~\text{or}~\mumu)+n\g$ may mimic our signal. The \FourS dataset contains a substantially larger admixture of such continuum backgrounds than the \ThreeS dataset as a result of the large integrated luminosity and relatively small production cross section for our signal. To reject such backgrounds in \FourS data, lepton candidates must have a value of $\mathcal{R}_e$ or $\mathcal{R}_\mu$ that exceeds 0.2. The identification efficiency for each muon~(electron) passing the likelihood ratio criterion is approximately $93\%$. Moreover, the lepton momenta must satisfy $p^\text{CM}<5.25~\gev$, to avoid a peak in continuum events near 5.29~\gev. To improve the purity in our search for $\cbj3 \to\om\OneS$, the electron mode is rejected with a selection of $\mathcal{R}_\mu>0.2$, and a more restrictive mass window of $M(\ellell)\in[9.2,9.6]~\gev$ is applied.\footnote{We use units in which the speed of light is $c=1$.}

Due to the limited phase space, all soft tracks in the CM frame, with $p^\text{CM}<0.43~\gev$ and $0.75~\gev$ for the $n=2$ and $n=3$ channels, respectively, are treated as pion candidates. Contamination from photon conversion to an \epem pair in detector material are suppressed by requiring that the cosine of the opening angle between oppositely-charged pion candidates be less than 0.95. To reject events with misreconstructed tracks, events containing multiple pairs of oppositely charged pions are rejected.

Photons are reconstructed from isolated clusters in the ECL that are not matched with a charged track projected from the CDC. To reject hadronic showers, the ratio of energy deposited in a $3\times3$ and $5\times5$ array of crystals centered on the most energetic one is required to exceed 90\%. Clusters with a transverse width exceeding $6~\cm$ are also rejected. To suppress beam-related backgrounds, photons are required to have an energy greater than $50~\mev$, $100~\mev$, and $150~\mev$ in the barrel, backward endcap, and forward endcap regions, respectively.

Neutral pion candidates are formed from combinations of photon pairs that satisfy $M(\g\g)\in[0.11,0.15]~\gev$. To reject combinatorial background from mis-reconstructed \pizz candidates, we require that $p^\text{CM}\in[0.08,0.43]~\gev$. A kinematic fit is performed to constrain the invariant mass of each candidate to the nominal \pizz mass \cite{pdg}, and the best-candidate \pizz is selected according the smallest mass-constrained fit $\chi^2$. Studies in simulation indicate that the best-candidate selection rejects 45\% of the background from misreconstructed \pizz at a cost of 14\% of the signal. The \om candidate is reconstructed as the combination of the \pizz and the \pipi pair, satisfying $\mom\in[0.71,0.83]~\gev$.

Backgrounds from resonant bottomonium di-pion transitions may mimic our final state $2\g2\pi2\ell$. The largest source of background is from $\TwoS\to\pipi\OneS$, which may be produced through feed-down decays $(\pipi,\pizz\pizz,\g\g~\text{via}~\cbj2 )$ of the \ThreeS or directly via ISR. In the \FourS dataset, additional contamination arises from transitions of the form $\FourS\to\pipi\TwoS$, where the \TwoS decays inclusively to the \OneS. To veto these backgrounds, we define a shifted mass difference $\dmpp = M(\pipi\ellell) - M(\ellell) + M(\OneS)$, where the broad resolution of the di-lepton invariant mass is removed by subtracting the reconstructed mass of the leptons and adding back the nominal \OneS mass \cite{pdg}. The di-pion transitions between \bbbar states give rise to narrow peaks in the \dmpp distribution with a resolution of approximately $2~\mev$.

Backgrounds from $\ThreeS\to\pipi\TwoS$ are rejected with $\dmpp > 9.83~\gev$, and pollution from $\ThreeS\to\pipi\OneS$ events is suppressed with $\dmpp<10.32~\gev$. Conveniently, the $\TwoS\to\pi\pi\OneS$ and $\FourS\to\pipi\TwoS$ backgrounds nearly overlap as $M(\FourS)-M(\TwoS) \approx M(\TwoS) - M(\OneS)$. The FOM optimization yields $\dmpp\notin(10.017,10.290)~\gev$ for the \ThreeS dataset and $\dmpp\notin(10.014,10.030)~\gev$ for the \FourS dataset. 

Table \ref{tab:Efficiencies} summarizes the selection efficiency for each channel. The \cbj3 efficiency is markedly lower due to the more restrictive requirements applied to the leptons.

\begin{table}[htb]
    \centering
    \caption{Selection efficiencies for each transition studied in large samples of MC simulated events.}
    \begin{tabular*}{0.35\textwidth}{c@{\extracolsep{\fill}}c}
        \hline\hline
        Transition & Efficiency (\%)\\
        \hline
        $\chibZero2 \to\om\OneS$ & $8.36\pm0.01$ \\
        $\chibOne2 \to\om\OneS$ &  $8.58\pm0.01$ \\
        $\chibTwo2 \to\om\OneS$ &  $8.59\pm0.01$ \\
        $\chibOne3 \to\om\OneS$ &  $5.37\pm0.02$ \\
        \hline\hline
    \end{tabular*}
    \label{tab:Efficiencies}
\end{table}

\section{Signal Extraction} \label{sec:signal_extraction}

\begin{figure}[htb]
    \begin{center}
    \includegraphics[width=0.45\textwidth]{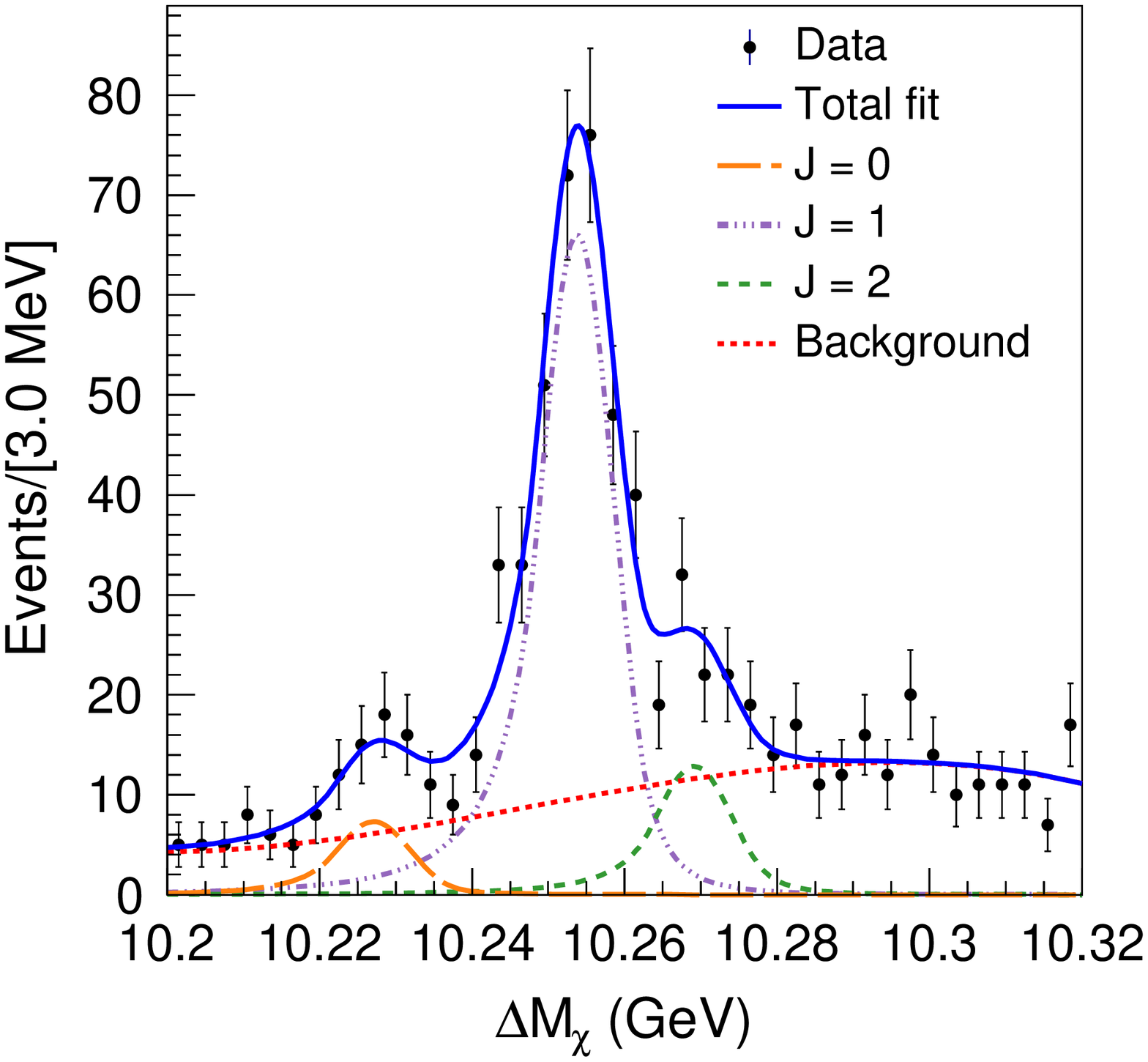}
    \qquad
    \includegraphics[width=0.45\textwidth]{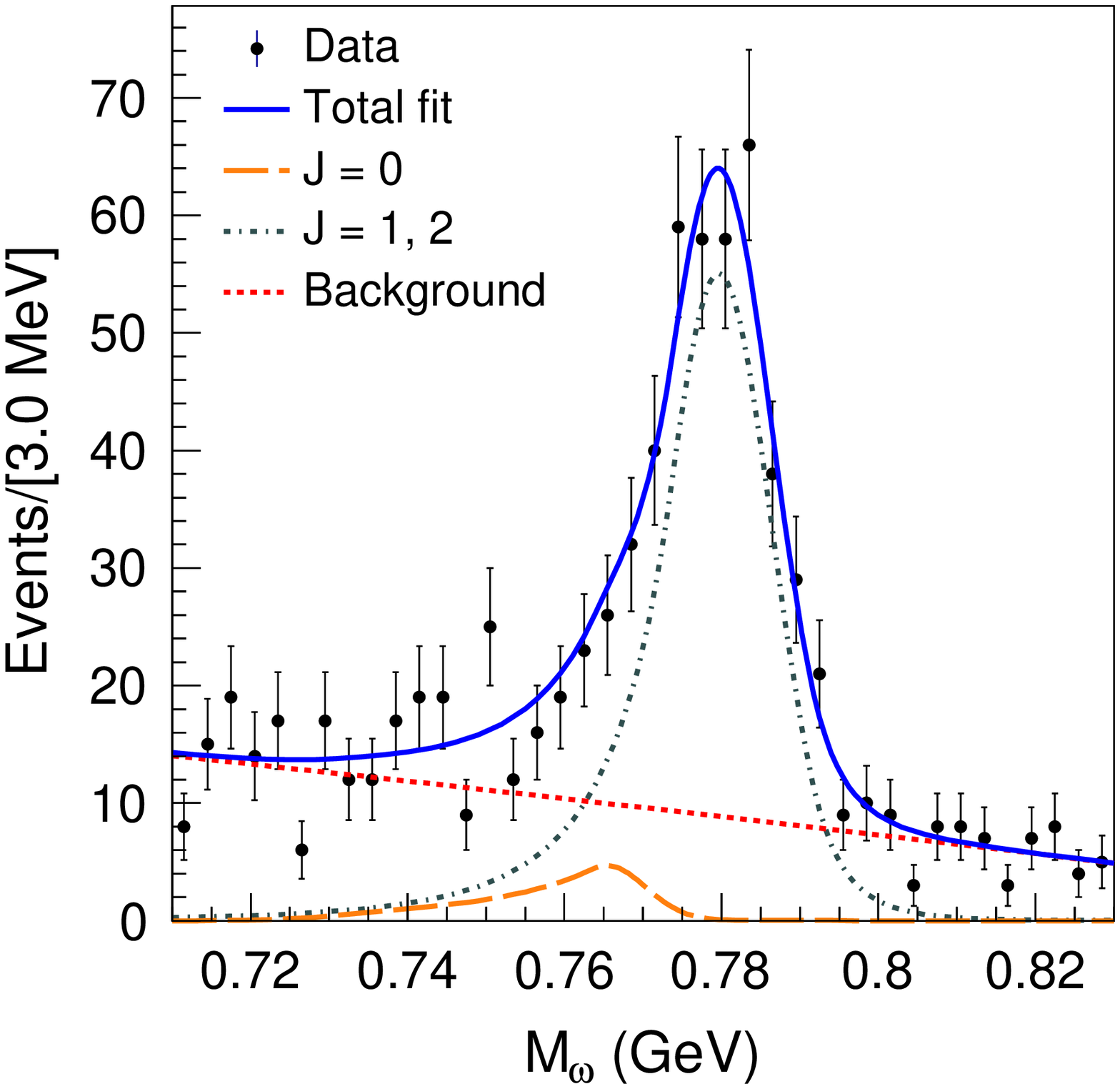}
    \caption{Fit to the \dmchi (Left) and \mom (Right) distributions for $\cbj2 \to\om\OneS$ candidates reconstructed in data. The solid blue curve shows the total fit and the dotted red curve indicates the background. In both panels, the long dashed orange curve is the \j0 signal. In the left panel, the dash-dotted violet curve is the \j1 signal, and the dashed green curve is the \j2 signal. In the right panel, the dash-dotted gray curve shows the combined \j1 and 2 signal.}
    \label{fig:2P_dataFit}
    \end{center}
\end{figure}

To discriminate amongst the \cbn signals, we define the shifted mass difference
\begin{equation}
    \dmchi = M(\pizz\pipi\ellell) - M(\ellell) + M(\OneS),
\end{equation}
where $M(\pizz\pipi\ellell)$ is the invariant mass of the final state, $M(\ellell)$ is the reconstructed $\Upsilon$ mass, and $M(\Upsilon)$ is the nominal mass from Ref. \cite{pdg}. The distribution of signal events is narrowly peaked at the corresponding \cbn mass, with a corresponding resolution of $4.5-6.5\mev$, depending on the transition. Signal yields are extracted from a simultaneous unbinned extended maximum-likelihood fit to the \dmchi and \om mass $(\mom)$ distributions. The projections of this fit are illustrated in Fig. \ref{fig:2P_dataFit}. The extracted signal yields are summarized in Table \ref{tab:yields}. 

All signal shapes, with the exception of the \mom component of the \j0 signal, are described by double-sided Crystal Ball (DSCB) functions \cite{DSCB}, which consist of a Gaussian core complemented by power-law tails on either side. The \j0 lineshape in \mom is impacted by the proximity of the $\om\OneS$ kinematic threshold, and so is parameterized as the product of a sigmoid and a DSCB function. Shape parameters are studied in simulation and fixed to the values extracted from fits to MC samples. In the fit to data, the means of the DSCB shapes are allowed to float along with multiplicative resolution calibrations defined independently for the \dmchi and \mom lineshapes. The backgrounds are modeled by cubic and quadratic functions in \dmchi and \mom, respectively. In the fit to data, the linear and quadratic coefficients are floated.

The statistical significance of each signal hypothesis is calculated using the profile likelihood method \cite{profLikeMethod}, and is summarized in Table \ref{tab:yields}. A fluctuation in excess of $3.2\sigma$ is observed that is consistent with the \j0 hypothesis, constituting the first evidence for a sub-threshold transition $\chibZero2 \to\om\OneS$.

\begin{table}[htb]
    \centering
    \caption{Extracted signal yields for various transitions and the associated significances, including systematic uncertainties, expressed in terms of standard deviations $(\sigma)$.}
    \begin{tabular*}{0.4\textwidth}{c@{\extracolsep{\fill}}cc}
        \hline\hline
        Transition & Signal Yield & Significance \\
        \hline
        $\chibZero2 \to\om\OneS$ & $33.1\asy{11.1}{10.8}$ & $3.2\sigma$\\
        $\chibOne2 \to\om\OneS$ & $309\pm24$ & $15.0\sigma$\\
        $\chibTwo2 \to\om\OneS$ & $62\pm16$ & $3.9\sigma$\\
        $\chibOne3 \to\om\OneS$ &  $3.2\asy{3.6}{2.8}$ & $1.1\sigma$\\
        \hline\hline
    \end{tabular*}
    \label{tab:yields}
\end{table}

\begin{figure}[htb]
    \centering
    \includegraphics[width=0.5\textwidth]{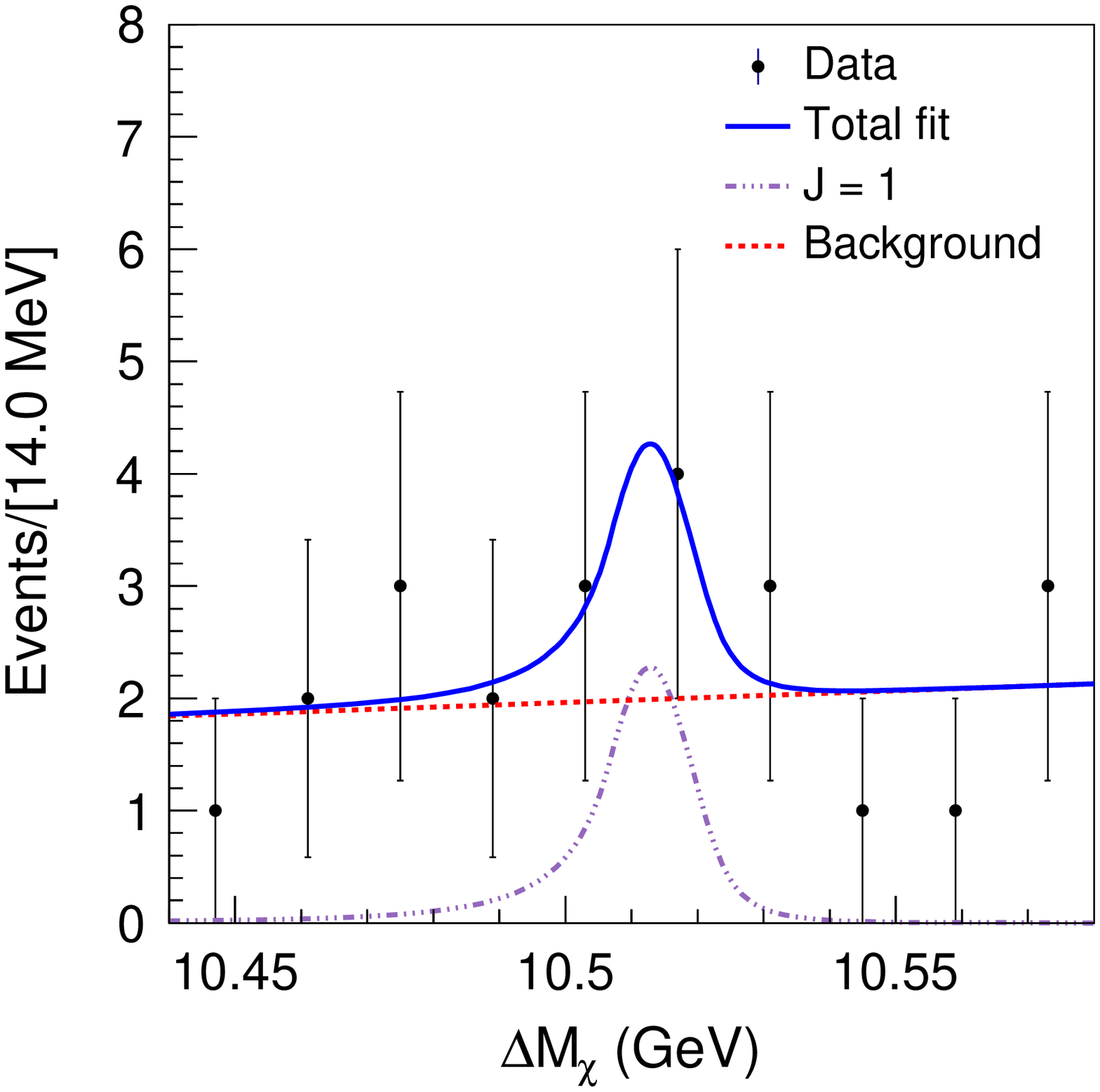}
    \caption{Fit to the \dmchi distribution for $\chibOne3 \to \om \OneS$ candidates reconstructed in data. The legend is similar to that of Fig. \ref{fig:2P_dataFit}.}
    \label{fig:3P_dataFit}
\end{figure}

Using the radiative branching fractions from Ref. \cite{Godfrey:2015} and estimating the branching fractions for $\cbj3 \to \om\OneS$, we project few signal events in data. Indeed, no \j0 or \j2 events are anticipated. As the \chibOne3 is expected to have the largest product branching fraction $\mathcal{B}\left(\FourS\to\g\chibOne3 \to\g\om\OneS\right)$, only the \j1 signal component is included in the fit to data. With a small number of signal events, the largest source of irreducible background arises from QED continuum events. This background is studied in off-resonance \FourS data, and modeled in the fit with a linear function. To stabilize the fit, the nominal \chibOne3 mass is fixed from Refs. \cite{pdg,cms3p}, and the calibration in the overall mass scale and resolution are determined from the control channel $\ThreeS\to\pipi\OneS$. From the fit to data, shown in Fig. \ref{fig:3P_dataFit}, we obtain a signal yield of $3.2\asy{3.6}{2.8}$ events.

\section{Systematic Uncertainties}

The uncertainties in the $\om\to\pipi\pizz$ and $\pizz\to\g\g$ external branching factions are treated as systematic uncertainties for all measurements. Additionally, the uncertainties in $\mathcal{B}\left(\ThreeS\to\g\cbj2 \right)$ and $\mathcal{B}\left(\ThreeS\to\pipi\OneS \right)$ impact the $2P$ measurements, and $\mathcal{B}\left(\OneS\to\mumu\right)$ affects the $3P$ measurement. The calculation of the $2P$ branching fractions relies on the measured $\ThreeS\to\pipi\OneS$ signal yield $(N_{\pi\pi\Upsilon})$. The uncertainty in the number of \ThreeS events is determined as the sum in quadrature of the statistical uncertainty and the systematic uncertainty ascribed for the fit procedure. The precision of the number of \FourS events, which is used in the calculation of the $3P$ branching fractions, is considered as an additional uncertainty.

Systematic uncertainties affecting the \cbj3 measurement, which cancel in the calculation of the \cbj2 branching fractions, include those assessed for data-MC differences in tracking and particle identification. A momentum-dependent systematic uncertainty for \pizz reconstruction is assessed and included.  Furthermore, a small contribution to the overall systematic uncertainty arises from MC statistics. 

The uncertainty due to the signal extraction procedure is estimated as the sum in quadrature of the results from the following studies. To estimate the impact of the choice of fit window and background parameterization, the fit to data is repeated with alternate fit windows in \dmchi and \mom as well as with nominal and alternate polynomials of higher order; the standard deviation of the resulting signal yields is used as an estimate of the systematic uncertainty. The impact of fixing the shape parameters from MC events is studied by repeating the fit to data while varying the fixed parameters according to the global covariance matrix; the standard deviation of the resulting distribution of signal yields is assessed as the systematic uncertainty. Finally, we search for bias in the fit with a set of toy MC studies with varied signal yields. The results form the basis of a linearity test from which we derive a small correction to the observed signal yields in data. The uncertainty for fit bias is assessed as half the relative difference between the corrected and nominal results. 

These uncertainties are combined in quadrature to obtain the total systematic uncertainty on each measurement. Table \ref{tab:systematics} summarizes the contribution of these sources of systematic uncertainty.

\begin{table}[htb]
    \centering
    \caption{Summary of systematic efficiencies impacting the branching fraction measurements, reported in percent. All sources except the signal extraction and selection efficiency cancel in the calculation of the ratios in Eq. \ref{eqn:rJ1}.}
    
    \begin{tabular*}{\textwidth}{l@{\extracolsep{\fill}}cccc}
        \hline\hline
        Source & $\mathcal{B}\left(\chibZero2 \to\om\Upsilon\right)$ & $\mathcal{B}\left(\chibOne2 \to\om\Upsilon\right)$ & $\mathcal{B}\left(\chibTwo2 \to\om\Upsilon\right)$ & 
        $\mathcal{B}\left(\FourS\to\g\chibOne3 \to\g\om\Upsilon\right)$ \\
        \hline
        Tracking & ... & ... & ... & ~$\pm1.4$ \\
        Particle Identification & ... & ... & ... & ~$\pm1.1$ \\
        \pizz reconstruction & ~$\pm1.7$ & ~$\pm1.7$ & ~$\pm1.7$ & ~$\pm3.3$ \\
        Selection Efficiency & ~$\pm0.1$ & ~$\pm0.1$ & ~$\pm0.1$ & ~~$\pm0.02$ \\
        Signal Extraction & ~\asy{8.7}{8.8} & ~\asy{1.1}{2.6} & ~\asy{3.6}{7.9} & \asy{10.1}{12.6}\\
        Number of \FourS & ... & ... & ... & ~$\pm1.4$ \\
        Number of \ThreeS & ~\asy{1.2}{1.1} & ~\asy{1.2}{1.1} & ~\asy{1.2}{1.1} & ... \\
        External Branching Fractions & $\pm10.4$ & ~$\pm9.4$ & $\pm12.4$ & ~$\pm2.2$ \\
        \hline
        Total & \asy{14.1}{14.2} & \asy{9.7}{10.0} & \asy{13.1}{14.8} & \asy{11.1}{13.4}\\
        \hline\hline
    \end{tabular*}
    \label{tab:systematics}
\end{table}

\section{Results}

With no significant \cbj3 signal observed, the \cbj2 reconstructed in \FourS data are attributed to radiative decays of \ThreeS mesons produced via ISR. The branching fractions for the \om transition are calculated from the signal yield $(N_J)$ and efficiency $(\epsilon_J)$ as
\begin{equation}
  \label{eqn:2P_bfCalculation}
%   \mathcal{B}(\chi_{bJ}(2P)\to\om\OneS) = \frac{N_{\chi J}}{\epsilon_{J} N_{\ThreeS} \Pi\\
    \mathcal{B}\left(\chi_{bJ}(2P)\to\om\OneS\right) = \frac{N_{J}}{N_{\pi\pi\Upsilon}}\frac{\epsilon_{\pi\pi\Upsilon}}{\epsilon_{J}}\frac{\mathcal{B}\big(\ThreeS\to\pipi\OneS\big)}{\mathcal{B}\big(\ThreeS\to\g\cbj2  \big)\mathcal{B}\big(\om\to\pipi\pizz\big)\mathcal{B}\big(\pizz\to\g\g\big)},     
\end{equation}
which incorporates the results of Appendix \ref{app:3S_count} for the number of \ThreeS events. In the ratio, the branching fraction of $\OneS\to\ellell$ drops out and several systematic uncertainties cancel. The resulting branching fractions are reported in Table \ref{tab:bfs}. These \j1,2 measurements are consistent within $2\sigma$ with the CLEO results \cite{toddOmega}.

We also reparameterize the fit in terms of the total signal yield and the ratios $\text{P}_{0/1}$ and $\text{P}_{2/1}$ between the \j0, 1 and \j2, 1 yields, respectively. Correcting the results for the efficiencies, we obtain the values of $r_{J/1}~=~\text{P}_{J/1}\left(\epsilon_{1}/\epsilon_{J}\right)$ shown in Table \ref{tab:bfs}. In each ratio $r_{J/1}$, only the systematic uncertainties assigned for signal extraction and the selection efficiency (on each yield) contribute. 

\begin{table}[htb]
    \centering
    \caption{Measured branching fractions (or upper limits) measured for each transition. The branching ratios $r_{0/1}$ and $r_{2/1}$ are also presented. The quoted uncertainties are statistical and systematic.}
    \begin{tabular*}{0.8\textwidth}{l@{\extracolsep{\fill}}ll}
        \hline\hline
        Quantity & Measurement (\%) & 90\% CL UL (\%) \\
        \hline
        $\mathcal{B}\left(\chibZero2 \to\om\OneS\right)$ & $0.56\asy{0.19}{0.18}\pm0.08$ & \\
        $\mathcal{B}\left(\chibOne2 \to\om\OneS\right)$ & $2.38\pm0.18\asy{0.23}{0.24}$ & \\
        $\mathcal{B}\left(\chibTwo2 \to\om\OneS\right)$ & $0.46\pm0.12\asy{0.06}{0.07}$ & \\
        $r_{0/1}$ & $0.110\asy{0.037}{0.036}\pm0.010$ & \\
        $r_{2/1}$ & $0.200\asy{0.062}{0.058}{}\asy{0.007}{0.017}$ & \\
        $\mathcal{B}\left(\FourS\to\g\chibOne3 \to\g\om\OneS\right)$ & $\left(4.9 \asy{5.5}{4.3}{}\asy{0.5}{0.6}\right)\times10^{-4}$ & $<1.4\times10^{-3}$ \\
        \hline\hline
    \end{tabular*}
    \label{tab:bfs}
\end{table}

We compare our measurement of $r_{2/1}$ with the QCDME expectation\cite{Voloshin:omegaDecayComments}, which we have calculated in Appendix \ref{app:r21_derivation} using current world averages\cite{pdg}: $r_{2/1}^\text{QCDME} = 0.77\pm0.16$. This reveals a tension with QCDME at the $3.3\sigma$ level.

We have also searched for the transition $\cbj3 \to\om\OneS$ produced in radiative decays of the \FourS. The branching fraction of the cascade transition is calculated as 
\begin{equation}
    \mathcal{B}\left(\FourS\to\g\chi_{bJ}(3P)\to\g\om\OneS\right) = \frac{N}{\epsilon N_{\FourS} \mathcal{B}\big(\om\to\pipi\pizz\big)\mathcal{B}\big(\pizz\to\g\g\big)\mathcal{B}\left(\OneS\to\mumu\right)},     
\end{equation}
where $N$ is the signal yield extracted from the fit to data, $\epsilon$ is the \chibOne3 selection efficiency, and $N_{\FourS}$ is the number of \FourS events. The result is presented in Table \ref{tab:bfs}. We obtain an upper limit on the cascade branching fraction by convolving the profile likelihood with a Gaussian function whose width equals systematic uncertainty and integrating over positive values of the branching fraction.  The result is an upper limit of $1.4\times10^{-5}$ set at 90\% confidence level (CL).

\section{Conclusions}

In summary, we have used the combined \ThreeS and \FourS data samples collected by the Belle detector to obtain first evidence for the near-threshold transition $\chibZero2 \to\om\OneS$ produced in radiative \ThreeS decays with a branching fraction of $(0.56\asy{0.18}{0.19}{}\asy{0.06}{0.07})\%$ at a significance of $3.2\sigma$. Moreover, we measure the hadronic transitions $\mathcal{B}\left(\chibOne2 \to\om\OneS\right) = (2.38\pm0.19\asy{0.23}{0.24})\%$  and $\mathcal{B}\left(\chibTwo2 \to\om\OneS\right) = (0.46\pm0.12\asy{0.06}{0.07})\%$. This constitutes the first confirmation of the \j1 and 2 branching fractions since their discovery \cite{toddOmega}. The ratios of the cascade branching fractions $(r_{J/1})$ are also measured. Comparison of the resulting measurement of $r_{2/1}$ with the value from QCDME reveals a $3.3\sigma$ tension. Finally, we search for $\cbj3 \to\om\OneS$ produced in radiative decays of the \FourS. As no significant signal is found, we set an upper limit on the cascade branching fraction $\mathcal{B}\left(\FourS\to\g\chibOne3 \to\g\om\OneS\right)<1.4\times10^{-5}$ at $90\%$ CL. 

\vspace{0.3cm}

\section{Acknowledgments}
\input{acknowledgments}

\appendix
\begin{appendices}

\section{Measurement of $\bm{N_{\ThreeS}}$} \label{app:3S_count}

The number of \ThreeS events is determined from the decay $\ThreeS\to\pipi\OneS$, with $\OneS\to\ellell$. The topology of this di-pion decay is well understood, and can be simulated with high fidelity using the measurements from Ref. \cite{cleoDipionAngularAnalysis}. This channel is used as a normalization channel for the $\cbj2 \to\pipi\OneS$ branching fraction measurement. 

Leptons and pions are reconstructed and combined to form di-pion and \OneS candidates with the event selection criteria specified in Sec. \ref{sec:event_selection}. Additionally, we require that the four charged tracks combine to form a shifted invariant mass $(\dmpp)$ within the signal region $\dmpp\in[10.32,10.39]~\gev$. Converted photons are vetoed by requiring that the cosine of the opening angle between charged pions be less than 0.95. Events containing multiple di-pion or \OneS candidates are rejected. The resulting reconstruction efficiency $(\epsilon_{\pi\pi\Upsilon})$ is $(41.41\pm0.02)\%$. 

\begin{figure}
    \centering
    \includegraphics[width=0.5\textwidth]{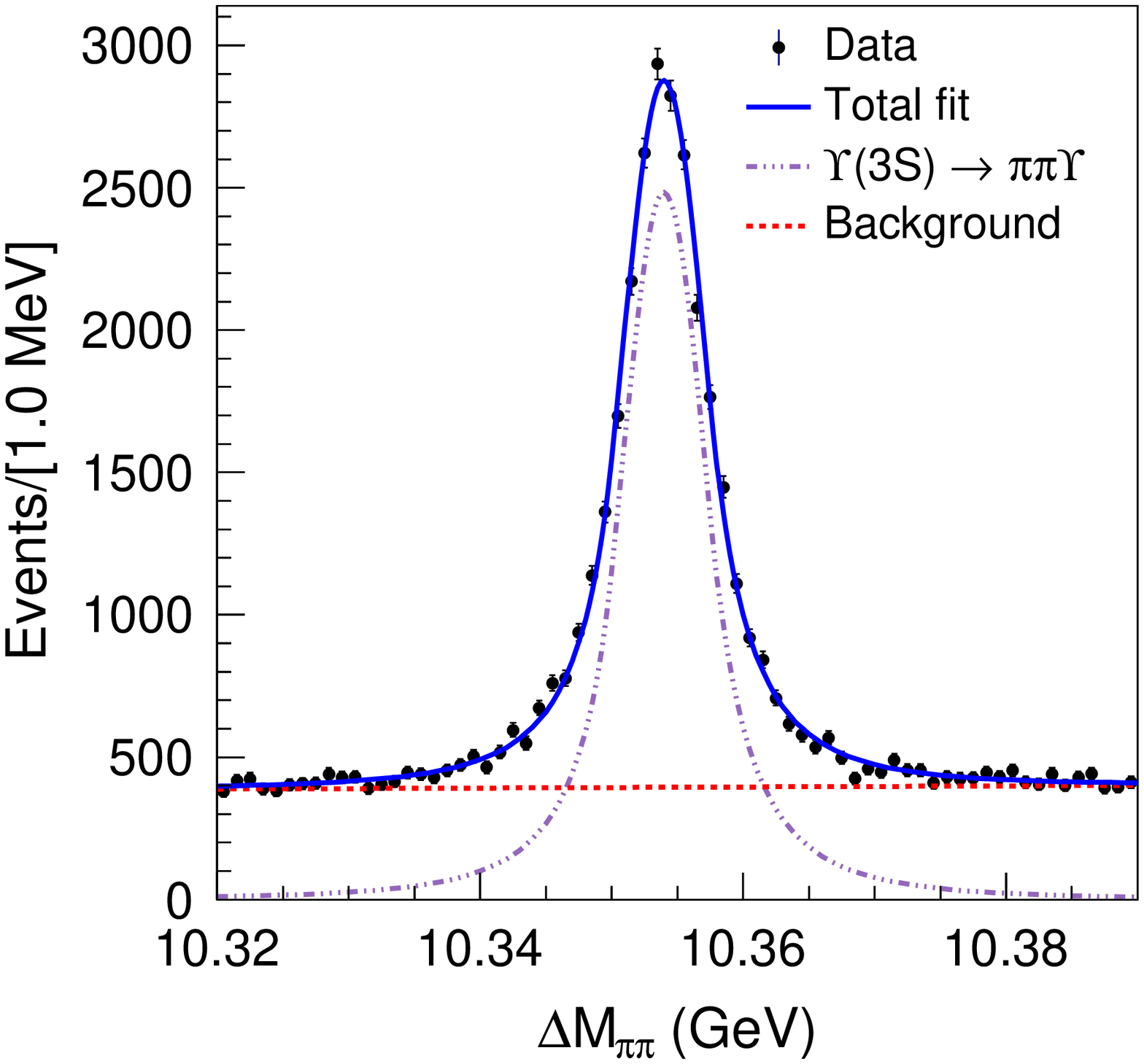}
    \caption{Fit to the \dmpp distribution for $\ThreeS \to\pipi\OneS$ candidates reconstructed in data. The total fit is shown by the solid blue curve, the background contribution by the dotted red curve, and the signal contribution by the dash-dotted violet curve is the overlaid signal shape.}
    \label{fig:dipi_dataFit}
\end{figure}

The signal yield is extracted from a fit to the \dmpp distribution shown in Fig. \ref{fig:dipi_dataFit}. The signal is parameterized with a DSCB and the background is described by a linear function. The observed signal yield is $N_{\pi\pi\Upsilon} = 24634\asy{228}{227}$. The number of \ThreeS candidates is calculated as:
\begin{equation}
    \label{eqn:N3S}
    N_{\ThreeS} = \frac{N_{\pi\pi\Upsilon}}{\epsilon_{\pi\pi\Upsilon}}\frac{1}{\mathcal{B}\big(\ThreeS\to\pipi\OneS\big)\mathcal{B}\big(\OneS\to\ellell\big)}.
\end{equation}
This corresponds to $(28.0\pm0.3\pm1.0)\times10^6$ events that contain an \ThreeS meson; here, the first uncertainty is statistical and the second is systematic.

The systematic uncertainty quoted above is determined as the sum in quadrature of the following sources: tracking $(1.5\%)$, particle identification (0.6\%), fit procedure $(\asy{0.8\%}{0.7\%})$, external branching fractions $(3.2\%)$, and the binomial uncertainty in the efficiency (0.02\%).

\section{Calculation of $\boldmath{r_{J/1}}$ from QCDME} \label{app:r21_derivation}

Following the discovery of $\cbj2 \to\om\OneS$, a derivation of $\boldmath{r_{2/1}^{QCDME}}$ from QCDME was published \cite{Voloshin:omegaDecayComments}. Using world average values from 2003, the ratio was calculated as $1.3\pm0.3$. In this appendix, the ratio is calculated using the current world averages \cite{pdg}, which benefit from the detailed studies of the radiative \bbbar transitions performed by BaBar \cite{BaBar:radCascade1,BaBar:radCascade2}.

In the ratio of the \j2 and 1 cascade transitions $\ThreeS\to\g\cbj2 \to\g\om\OneS$, the total width of the \ThreeS cancels:
    \begin{equation}
        \begin{split}
            r_{2/1}^{QCDME} &= \frac{\Gamma(\ThreeS \to\g\chibTwo2 )}{\Gamma(\ThreeS\to\g\chibOne2 )}\frac{\Gamma(\chibTwo2 \to \om\OneS)}{\Gamma(\chibOne2 \to \om\OneS)}\frac{\Gamma(\chibOne2 )}{\Gamma(\chibTwo2 )}\\
            &\equiv G_{2/1} \times R_{2/1} \times W_{1/2},\\
        \end{split}
        \label{eqn:3s_r21}
    \end{equation}
where $G_{2/1}$ is the ratio of the \ThreeS radiative decay partial widths, $R_{2/1}$ is the ratio of the $\chi_{bJ}(2P) \to\om\OneS$ partial widths, and $W_{1/2}$ is the ratio of the \cbj2 total widths. 

The uncertainties in the measured \ThreeS radiative branching fractions are prohibitively large to use in the calculation of $G_{2/1}$. Instead, in keeping with Ref. \cite{Voloshin:omegaDecayComments}, the radiative decay partial widths of the \ThreeS are expanded using the dipole transition formula from QCDME
\begin{equation} \label{eqn:r21_radiativeWidthQCDME}
    \Gamma(\ThreeS\to\g\cb )\approx (2J+1)k_{\g}^3,
\end{equation}
where $J$ is the total angular momentum of the final state and $k_{\g}$ is the well-measured photon momentum \cite{pdg}. Employing Eq. \ref{eqn:r21_radiativeWidthQCDME}, we calculate $G_{2/1} = 1.091 \pm 0.011$.

In the nonrelativistic limit of QCDME, the spin dependence of the decay amplitude factorizes. In the ratio $R_{2/1}$, the spin of the heavy quark decouples and the quantity may be approximated as the ratio of the $S$-wave phase space factors $(\Delta_i)$:
\begin{equation} \label{eqn:r21_R21}
    \begin{split}
      R_{2/1} &= \sqrt{\frac{\Delta_2}{\Delta_1}} = \sqrt{1+\frac{\Delta M_{2/1}}{\Delta_1}},
    \end{split}
\end{equation}
where $\Delta_1 = M\left(\chibOne2 \right) - M\left(\OneS\right) - M\left(\om\right)$, and $\Delta M_{2/1}$ is the mass splitting between the \j1 and 2 states. We obtain $R_{2/1} = 1.431\pm0.019$.

The ratio of the \cb total widths $(W_{1/2})$ is determined by expanding the ratio of the \cbj2 radiative transitions to lower $\Upsilon(mS)$ states using dipole transition formulae, analogous to Eq. \ref{eqn:r21_radiativeWidthQCDME}, and noting that \j1 for $\Upsilon(mS)$ states:
\begin{equation} \label{eqn:r21_W12}
    \begin{split}
        \frac{\mathcal{B}(\chibTwo2 \to \g\Upsilon(mS))}{\mathcal{B}(\chibOne2 \to \g\Upsilon(mS))} &= \frac{\Gamma(\chibTwo2 \to \g\Upsilon(mS))}{\Gamma(\chibOne2 \to \g\Upsilon(mS))}\frac{\Gamma(\chibOne2 )}{\Gamma(\chibTwo2 )}\\
        &= \left[\frac{k_{\gamma}\left(\chibTwo2  \to\g\Upsilon(mS)\right)}{k_{\gamma}\left(\chibOne2 \to\g\Upsilon(mS)\right)}\right]^3 W_{1/2}.
    \end{split}
\end{equation}
Rewriting the photon momenta in terms of the various masses and solving for $W_{1/2}$ utilizing the radiative branching fractions from Ref. \cite{pdg}, we obtain:
\begin{equation} \label{eqn:r21_W12_vals}
    W_{1/2} =
    \begin{cases}
        \begin{array}{lr}
            0.6359\pm0.1003, & m=1\\
            0.4179\pm0.0714, & m=2
        \end{array}
    \end{cases}
\end{equation}
These values should be compared with the value $W_{1/2} = 0.80\pm0.15$ from Ref. \cite{Voloshin:omegaDecayComments}, which used only the $m=1$ transitions, citing prohibitively large uncertainties in the measurements of the $m=2$ channel. We form a least-squares weighted average\cite{pdg} of the values in Eq. \ref{eqn:r21_W12_vals} and obtain a value of $W_{1/2} = 0.49\pm0.10$, where the uncertainty has been inflated by a factor of $S = \sqrt{\chi^2 / (N - 1) } = 1.77$. We obtain the recalculated ratio $r_{2/1} = 0.77 \pm 0.16$.
\end{appendices}

\bibliography{belle}

\end{document}

%% file: authors.tex
%%% Paper:
%%% Journal:  2021 Conference Papers
%%% February 10, 2021 - first edition (from last edition of author-conf2020.tex)
%%% February 10, 2021 - Liptak affiliation
%%% March 5, 2021 - add Gong
%%% April 16, 2021 - Grzymkowska->Werbycka
%%% April 22, 2021 - Sato affiliation
%%% April 30, 2021 - add T.Gu, Ecker, Tiwary
%%% May 14, 2021 - add RomaTre: Branchini, Budano, De Pietro, Graziani, Laurenza, Passeri
%%% June 1, 2021 - Lai affiliation
%%% June 13, 2021 - Hernandez-Villanueva affiliation
%%% June 14, 2021 - add D.Wang
%%% July 12, 2021 - add Bodrov
%%% Contacts:
%%% Non-responding authors or those who said NO are commented out.
%%% ====================================================================
%%% Click the RELOAD button on your web browser to see the updated file.
%%% ====================================================================
%%% Use \input{author} to insert this material into your latex file.
%%%%% Force institutions to appear in alphabetical order when typeset.
\noaffiliation
\affiliation{Department of Physics, University of the Basque Country UPV/EHU, 48080 Bilbao}
\affiliation{Beihang University, Beijing 100191}
\affiliation{University of Bonn, 53115 Bonn}
\affiliation{Brookhaven National Laboratory, Upton, New York 11973}
\affiliation{Budker Institute of Nuclear Physics SB RAS, Novosibirsk 630090}
\affiliation{Faculty of Mathematics and Physics, Charles University, 121 16 Prague}
\affiliation{Chiba University, Chiba 263-8522}
\affiliation{Chonnam National University, Gwangju 61186}
\affiliation{University of Cincinnati, Cincinnati, Ohio 45221}
\affiliation{Deutsches Elektronen--Synchrotron, 22607 Hamburg}
\affiliation{Duke University, Durham, North Carolina 27708}
\affiliation{University of Florida, Gainesville, Florida 32611}
\affiliation{Department of Physics, Fu Jen Catholic University, Taipei 24205}
\affiliation{Key Laboratory of Nuclear Physics and Ion-beam Application (MOE) and Institute of Modern Physics, Fudan University, Shanghai 200443}
\affiliation{Justus-Liebig-Universit\"at Gie\ss{}en, 35392 Gie\ss{}en}
\affiliation{Gifu University, Gifu 501-1193}
\affiliation{II. Physikalisches Institut, Georg-August-Universit\"at G\"ottingen, 37073 G\"ottingen}
\affiliation{SOKENDAI (The Graduate University for Advanced Studies), Hayama 240-0193}
\affiliation{Gyeongsang National University, Jinju 52828}
\affiliation{Department of Physics and Institute of Natural Sciences, Hanyang University, Seoul 04763}
\affiliation{University of Hawaii, Honolulu, Hawaii 96822}
\affiliation{High Energy Accelerator Research Organization (KEK), Tsukuba 305-0801}
\affiliation{J-PARC Branch, KEK Theory Center, High Energy Accelerator Research Organization (KEK), Tsukuba 305-0801}
\affiliation{National Research University Higher School of Economics, Moscow 101000}
\affiliation{Forschungszentrum J\"{u}lich, 52425 J\"{u}lich}
\affiliation{Hiroshima Institute of Technology, Hiroshima 731-5193}
\affiliation{Hiroshima University, Higashi-Hiroshima, Hiroshima 739-8530}
\affiliation{IKERBASQUE, Basque Foundation for Science, 48013 Bilbao}
\affiliation{University of Illinois at Urbana-Champaign, Urbana, Illinois 61801}
\affiliation{Indian Institute of Science Education and Research Mohali, SAS Nagar, 140306}
\affiliation{Indian Institute of Technology Bhubaneswar, Satya Nagar 751007}
\affiliation{Indian Institute of Technology Guwahati, Assam 781039}
\affiliation{Indian Institute of Technology Hyderabad, Telangana 502285}
\affiliation{Indian Institute of Technology Madras, Chennai 600036}
\affiliation{Indiana University, Bloomington, Indiana 47408}
\affiliation{Institute of High Energy Physics, Chinese Academy of Sciences, Beijing 100049}
\affiliation{Institute of High Energy Physics, Vienna 1050}
\affiliation{Institute for High Energy Physics, Protvino 142281}
\affiliation{Institute of Mathematical Sciences, Chennai 600113}
\affiliation{INFN - Sezione di Napoli, 80126 Napoli}
\affiliation{INFN - Sezione di Roma Tre, I-00146 Roma}
\affiliation{INFN - Sezione di Torino, 10125 Torino}
\affiliation{Advanced Science Research Center, Japan Atomic Energy Agency, Naka 319-1195}
\affiliation{J. Stefan Institute, 1000 Ljubljana}
\affiliation{Kanagawa University, Yokohama 221-8686}
\affiliation{Institut f\"ur Experimentelle Teilchenphysik, Karlsruher Institut f\"ur Technologie, 76131 Karlsruhe}
\affiliation{Kavli Institute for the Physics and Mathematics of the Universe (WPI), University of Tokyo, Kashiwa 277-8583}
\affiliation{Kennesaw State University, Kennesaw, Georgia 30144}
\affiliation{King Abdulaziz City for Science and Technology, Riyadh 11442}
\affiliation{Department of Physics, Faculty of Science, King Abdulaziz University, Jeddah 21589}
\affiliation{Kitasato University, Sagamihara 252-0373}
\affiliation{Korea Institute of Science and Technology Information, Daejeon 34141}
\affiliation{Korea University, Seoul 02841}
\affiliation{Kyoto Sangyo University, Kyoto 603-8555}
\affiliation{Kyoto University, Kyoto 606-8502}
\affiliation{Kyungpook National University, Daegu 41566}
\affiliation{Universit\'{e} Paris-Saclay, CNRS/IN2P3, IJCLab, 91405 Orsay}
\affiliation{\'Ecole Polytechnique F\'ed\'erale de Lausanne (EPFL), Lausanne 1015}
\affiliation{P.N. Lebedev Physical Institute of the Russian Academy of Sciences, Moscow 119991}
\affiliation{Liaoning Normal University, Dalian 116029}
\affiliation{Faculty of Mathematics and Physics, University of Ljubljana, 1000 Ljubljana}
\affiliation{Ludwig Maximilians University, 80539 Munich}
\affiliation{Luther College, Decorah, Iowa 52101}
\affiliation{Malaviya National Institute of Technology Jaipur, Jaipur 302017}
\affiliation{University of Malaya, 50603 Kuala Lumpur}
\affiliation{Faculty of Chemistry and Chemical Engineering, University of Maribor, 2000 Maribor}
\affiliation{Max-Planck-Institut f\"ur Physik, 80805 M\"unchen}
\affiliation{School of Physics, University of Melbourne, Victoria 3010}
\affiliation{University of Mississippi, University, Mississippi 38677}
\affiliation{University of Miyazaki, Miyazaki 889-2192}
\affiliation{Moscow Physical Engineering Institute, Moscow 115409}
\affiliation{Graduate School of Science, Nagoya University, Nagoya 464-8602}
\affiliation{Kobayashi-Maskawa Institute, Nagoya University, Nagoya 464-8602}
\affiliation{Universit\`{a} di Napoli Federico II, 80126 Napoli}
\affiliation{Nara University of Education, Nara 630-8528}
\affiliation{Nara Women's University, Nara 630-8506}
\affiliation{National Central University, Chung-li 32054}
\affiliation{National United University, Miao Li 36003}
\affiliation{Department of Physics, National Taiwan University, Taipei 10617}
\affiliation{H. Niewodniczanski Institute of Nuclear Physics, Krakow 31-342}
\affiliation{Nippon Dental University, Niigata 951-8580}
\affiliation{Niigata University, Niigata 950-2181}
\affiliation{University of Nova Gorica, 5000 Nova Gorica}
\affiliation{Novosibirsk State University, Novosibirsk 630090}
\affiliation{Okinawa Institute of Science and Technology, Okinawa 904-0495}
\affiliation{Osaka City University, Osaka 558-8585}
\affiliation{Osaka University, Osaka 565-0871}
\affiliation{Pacific Northwest National Laboratory, Richland, Washington 99352}
\affiliation{Panjab University, Chandigarh 160014}
\affiliation{Peking University, Beijing 100871}
\affiliation{University of Pittsburgh, Pittsburgh, Pennsylvania 15260}
\affiliation{Punjab Agricultural University, Ludhiana 141004}
\affiliation{Research Center for Electron Photon Science, Tohoku University, Sendai 980-8578}
\affiliation{Research Center for Nuclear Physics, Osaka University, Osaka 567-0047}
\affiliation{Meson Science Laboratory, Cluster for Pioneering Research, RIKEN, Saitama 351-0198}
\affiliation{Theoretical Research Division, Nishina Center, RIKEN, Saitama 351-0198}
\affiliation{RIKEN BNL Research Center, Upton, New York 11973}
\affiliation{Dipartimento di Matematica e Fisica, Universit\`{a} di Roma Tre, I-00146 Roma}
\affiliation{Saga University, Saga 840-8502}
\affiliation{Department of Modern Physics and State Key Laboratory of Particle Detection and Electronics, University of Science and Technology of China, Hefei 230026}
\affiliation{Seoul National University, Seoul 08826}
\affiliation{Showa Pharmaceutical University, Tokyo 194-8543}
\affiliation{Soochow University, Suzhou 215006}
\affiliation{Soongsil University, Seoul 06978}
\affiliation{University of South Carolina, Columbia, South Carolina 29208}
\affiliation{Stefan Meyer Institute for Subatomic Physics, Vienna 1090}
\affiliation{Sungkyunkwan University, Suwon 16419}
\affiliation{School of Physics, University of Sydney, New South Wales 2006}
\affiliation{Department of Physics, Faculty of Science, University of Tabuk, Tabuk 71451}
\affiliation{Tata Institute of Fundamental Research, Mumbai 400005}
\affiliation{Excellence Cluster Universe, Technische Universit\"at M\"unchen, 85748 Garching}
\affiliation{Department of Physics, Technische Universit\"at M\"unchen, 85748 Garching}
\affiliation{School of Physics and Astronomy, Tel Aviv University, Tel Aviv 69978}
\affiliation{Toho University, Funabashi 274-8510}
\affiliation{Department of Physics, Tohoku University, Sendai 980-8578}
\affiliation{Earthquake Research Institute, University of Tokyo, Tokyo 113-0032}
\affiliation{Department of Physics, University of Tokyo, Tokyo 113-0033}
\affiliation{Tokyo Institute of Technology, Tokyo 152-8550}
\affiliation{Tokyo Metropolitan University, Tokyo 192-0397}
\affiliation{Tokyo University of Agriculture and Technology, Tokyo 184-8588}
\affiliation{Utkal University, Bhubaneswar 751004}
\affiliation{Virginia Polytechnic Institute and State University, Blacksburg, Virginia 24061}
\affiliation{Wayne State University, Detroit, Michigan 48202}
\affiliation{Yamagata University, Yamagata 990-8560}
\affiliation{Yonsei University, Seoul 03722}
%   \author{Z.~S.~Stottler}\affiliation{Virginia Polytechnic Institute and State University, Blacksburg, Virginia 24061} % VPI
%   \author{T.~K.~Pedlar}\affiliation{Luther College, Decorah, Iowa 52101} % Luther
%   \author{B.~G.~Fulsom}\affiliation{Pacific Northwest National Laboratory, Richland, Washington 99352} % PNNL
%   \author{L.~E.~Piilonen}\affiliation{Virginia Polytechnic Institute and State University, Blacksburg, Virginia 24061} % VPI
  \author{A.~Abdesselam}\affiliation{Department of Physics, Faculty of Science, University of Tabuk, Tabuk 71451} % Tabuk
  \author{I.~Adachi}\affiliation{High Energy Accelerator Research Organization (KEK), Tsukuba 305-0801}\affiliation{SOKENDAI (The Graduate University for Advanced Studies), Hayama 240-0193} % KEK
  \author{K.~Adamczyk}\affiliation{H. Niewodniczanski Institute of Nuclear Physics, Krakow 31-342} % Krakow
  \author{J.~K.~Ahn}\affiliation{Korea University, Seoul 02841} % Korea
  \author{H.~Aihara}\affiliation{Department of Physics, University of Tokyo, Tokyo 113-0033} % Tokyo
  \author{S.~Al~Said}\affiliation{Department of Physics, Faculty of Science, University of Tabuk, Tabuk 71451}\affiliation{Department of Physics, Faculty of Science, King Abdulaziz University, Jeddah 21589} % Tabuk
  \author{K.~Arinstein}\affiliation{Budker Institute of Nuclear Physics SB RAS, Novosibirsk 630090}\affiliation{Novosibirsk State University, Novosibirsk 630090} % BINP
  \author{Y.~Arita}\affiliation{Graduate School of Science, Nagoya University, Nagoya 464-8602} % Nagoya
  \author{D.~M.~Asner}\affiliation{Brookhaven National Laboratory, Upton, New York 11973} % BNL
  \author{H.~Atmacan}\affiliation{University of Cincinnati, Cincinnati, Ohio 45221} % Cincinnati
% ...
% Lausanne
\collaboration{The Belle Collaboration}

%% file: acknowledgments.tex
We thank the KEKB group for the excellent operation of the
accelerator; the KEK cryogenics group for the efficient
operation of the solenoid; and the KEK computer group, and the Pacific Northwest National
Laboratory (PNNL) Environmental Molecular Sciences Laboratory (EMSL)
computing group for strong computing support; and the National
Institute of Informatics, and Science Information NETwork 5 (SINET5) for
valuable network support.  We acknowledge support from
the Ministry of Education, Culture, Sports, Science, and
Technology (MEXT) of Japan, the Japan Society for the 
Promotion of Science (JSPS), and the Tau-Lepton Physics 
Research Center of Nagoya University; 
the Australian Research Council including grants
DP180102629, % Sevior
DP170102389, % Varvell
DP170102204, % Yabsley
DP150103061, % Urquijo
FT130100303; % Urquijo;
Austrian Federal Ministry of Education, Science and Research (FWF) and
FWF Austrian Science Fund No.~P~31361-N36;
the National Natural Science Foundation of China under Contracts
No.~11435013,  %Zhen-An Liu
No.~11475187,  %Chang-Zheng Yuan
No.~11521505,  %Chang-Zheng Yuan
No.~11575017,  %Cheng-Ping Shen
No.~11675166,  %Wen-Biao Yan
No.~11705209;  %Yi-Ming Li
Key Research Program of Frontier Sciences, Chinese Academy of Sciences (CAS), Grant No.~QYZDJ-SSW-SLH011; % Chang-Zheng Yuan
the  CAS Center for Excellence in Particle Physics (CCEPP); %Chang-Zheng Yuan,
the Shanghai Science and Technology Committee (STCSM) under Grant No.~19ZR1403000; %Xiaolong Wang
the Ministry of Education, Youth and Sports of the Czech
Republic under Contract No.~LTT17020;
Horizon 2020 ERC Advanced Grant No.~884719 and ERC Starting Grant No.~947006 ``InterLeptons'' (European Union);
the Carl Zeiss Foundation, the Deutsche Forschungsgemeinschaft, the
Excellence Cluster Universe, and the VolkswagenStiftung;
the Department of Atomic Energy (Project Identification No. RTI 4002) and the Department of Science and Technology of India; 
the Istituto Nazionale di Fisica Nucleare of Italy; 
National Research Foundation (NRF) of Korea Grant
Nos.~2016R1\-D1A1B\-01010135, 2016R1\-D1A1B\-02012900, 2018R1\-A2B\-3003643,
2018R1\-A6A1A\-06024970, 2018R1\-D1A1B\-07047294, 2019K1\-A3A7A\-09033840,
2019R1\-I1A3A\-01058933;
Radiation Science Research Institute, Foreign Large-size Research Facility Application Supporting project, the Global Science Experimental Data Hub Center of the Korea Institute of Science and Technology Information and KREONET/GLORIAD;
the Polish Ministry of Science and Higher Education and 
the National Science Center;
the Ministry of Science and Higher Education of the Russian Federation, Agreement 14.W03.31.0026, % from 15.02.2018
and the HSE University Basic Research Program, Moscow; % from 15.04.2021
University of Tabuk research grants
S-1440-0321, S-0256-1438, and S-0280-1439 (Saudi Arabia);
the Slovenian Research Agency Grant Nos. J1-9124 and P1-0135;
Ikerbasque, Basque Foundation for Science, Spain;
the Swiss National Science Foundation; 
the Ministry of Education and the Ministry of Science and Technology of Taiwan;
and the United States Department of Energy and the National Science Foundation.